# *Voice Biomarker Identification for Effects of Deep-Brain Stimulation on Parkinson's Disease*

Huy Phi*, Sanjeev Janarthanan*, Larry Zhang*, Reza Hosseini Ghomi*

*DigiPsych Lab, Department of Neurology, University of Washington, Seattle, WA, USA

{huyphi, sanjrj, lazhang, rezahg}@uw.edu

**Abstract**

Deep-Brain Stimulation (DBS) is a therapy used in conjunction with medication to help alleviate the motor symptoms of Parkinson's Disease (PD). However, the monitoring and adjustment of DBS settings is tedious and expensive, requiring long programming appointments every few months. We investigated the possible correlation between PD motor score severity and digitally extracted patient voice features to potentially aid clinicians in their monitoring and treatment of PD with DBS to eventually enable a closed-loop DBS system. 5 DBS PD patients (4 bilateral STN, 1 bilateral GPI) were enrolled. Voice samples were collected for various voice tasks (single phoneme vocalization, free speech task, sentence reading task, counting backward task, categorical fluency task) for DBS in ON and OFF states. Motor scores per the Unified Parkinson's Disease Rating Scale (UPDRS) were also collected for DBS ON and OFF states. Voice samples were then analyzed to extract voice features using publicly available voice feature library sets, and statistically compared for DBS ON and OFF. Of the feature categories explored (Acoustic, Prosodic, Linguistic) 6 features from the GeMAPS feature set for acoustic features demonstrated significant differences with DBS ON and OFF ($p < 0.05$). Prosodic features such as pause length/percentage were found to be negatively correlated with increased motor symptom severity. Non-significant differences were found for linguistic features. These findings provide preliminary evidence for acoustic and prosodic speech features to act as potential biomarkers for PD disease severity in DBS patients. We hope to explore further by expanding our data set, identifying other features and applying potential machine learning models, working towards a closed-loop DBS system that can auto-tune itself based on changes in a patient's voice.

Keywords: Deep-Brain Stimulation, Parkinson's Disease, PD-Related Motor Symptoms, Digital Biomarkers, Voice Technology, Audio Features, Voice Biomarker.

## 1. Introduction:

*1.1 Parkinson's and Deep-Brain Stimulation:*

Parkinson's Disease (PD) is the second most common neurodegenerative disease, expected to affect 930,000 patients in the US by 2020, projecting to over 1.3 million people by 2030 [1]. Known for its motor symptoms of muscle rigidity, bradykinesia and tremor, PD also effects the fundamental frequency , jitter, and harshness of voice [2]. Along with medication, PD can be treated using Deep-Brain Stimulation (DBS), where an electrode is surgically implanted into the brain, delivering electric current to alleviate symptoms. Specific targets for DBS include the sub-thalamic nuclei (STN) or globus pallidus internus (GPI) in the indirect pathway of the basal ganglia thalamocortical circuit, the pathway contributing to the inhibition of movements associated with PD pathology [3]. The best DBS candidates for PD include patients with significant responses to levodopa, and lack of comorbidities, particularly cognitive and surgically related [4]. An estimated 150,000 patients with movement disorders (PD, essential tremor, etc.) have been treated with DBS in the United States [5].

*1.2 Voice Biomarker Closed-Loop DBS:*

DBS treatment is an arduous and expensive process. Patients must come in every few months to have their DBS settings carefully adjusted with the progressions of their symptoms. While modern DBS devices have wireless charging capabilities, older devices must have their batteries surgically replaced every few years. Closed-loop DBS refers to a system which makes adjustment to the DBS signal without the need for manual adjustment. This could reduce the need for patients to come into clinic, and prolong the battery life of the function generators on their DBS devices. Closed-loop DBS has been explored using physiological feedback loops, particularly with motor cortex electrocorticography (ECoG), electroencephalography (EEG), and wearable wrist accelerometer as physiological sensors [6]&[7]. Previously, our lab has identified methods to utilize digital biomarkers extracted from voice samples, to build models that can predict presence of PD [8]. Other work has shown changes in vocal motor mechanisms of PD with DBS ON and OFF [9], and changes in the quality, intensity and prosody of patient voice for PD amongst different DBS settings [10]. We aim to identify objective and quantifiable voice features that correlate with DBS turned ON and OFF in PD patients, that we could potentially apply to use as feedback for closed-loop system DBS.

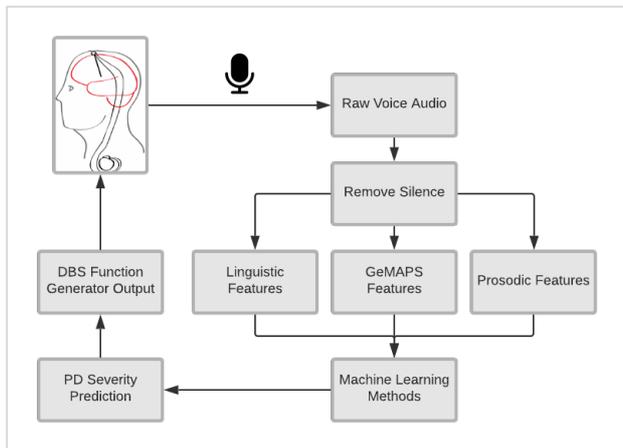

Figure 1: Proposed Closed-Loop DBS System

## 2. Materials and Methods:

*2.1 Data Collection:*

Selection criteria for this study were any patients who underwent DBS treatment for PD at the University of Washington Medical Center. 5 DBS PD patients were enrolled in this study (3 male, 2 female). Patient demographics are displayed on Table 1. 4 of the patients underwent bilateral STN targeted DBS, and 1 patient underwent bilateral GPI targeted DBS (patient 5*). Levodopa Equivalency Dosage (LED) was calculated per Tomlinson et al. [11], and is used to normalize patient dosages to a Levodopa equivalent.

*2.2 Motor Scores:*

Motor scores per the Movement Disorder Society Unified Parkinson's Disease Rating Scale Part III (UPDRS) [12] were collected for each patient for DBS in the ON and OFF states by a trained movement disorders neurologist. The UPDRS part III is a rating tool used to determine the severity of motor symptoms (e.g. muscle rigidity, posture, finger tapping, tremor etc.) Each item is rated on a scale of 0-4, with 4 being the most severe. The UPDRS Part III has a total of 33 items for a total score range of 0-132. Scores recorded were used to demonstrate the change in PD motor symptom severity for DBS ON and OFF.

*2.3 Voice Tasks:*

Voice recordings were collected using smartphone microphone, located an arm's length away from the mouth. 5 distinct voice tasks for DBS ON and OFF were collected, for a total of 10 voice recordings for each patient. These voice tasks included 10 seconds of single phoneme sound, /ˈɑ/, pronounced "ahhh", 30 second free speech task, standardized sentence reading task, counting backwards task, and a 30 second categorical fluency task – having participants name as many animals as they could in the allotted time. Transcripts for free speech and categorical fluency tasks were also generated for text analysis.

*2.4 Feature Extraction:*

Voice samples were extracted for features using publicly available feature extraction pipelines [13] and [14]. Acoustic features extracted were derived from the Geneva Minimalistic Acoustic Parameter Set (GeMAPS), consisting of 88 acoustic voice features, such as fundamental frequency, jitter, shimmer, etc. [15]. The second feature set extracted was for prosody features, which included 7 features looking at speech/pause characteristics [16]. Patient free speech tasks were used to extract Linguistic Features of Complexity set, including 7 features identifying lexical richness [16], and Natural Language Toolkit (NLTK) set, including 37 features identifying linguistic repeats, consonant, and vowel use [17].

*2.5 Data Analysis:*

To identify significant changes for DBS ON compared to off, paired T-tests were performed on all extracted features. Pearson correlation coefficients were generated to identify correlation between prosody features and motor symptom severity.

| Patient ID | Age/Sex | PD Duration (Year) | Time from DBS Surgery (Year) | UPDRS-III Scores OFF/ON | Levodopa Equivalent Dosage |
|---|---|---|---|---|---|
| 1 | 86/F | 5 | 0.83 | 36/30 | 75 |
| 2 | 69/M | 15 | 0.83 | 23/15 | 2000 |
| 3 | 71/F | 6 | 0.83 | 27/21 | 800 |
| 4 | 52/M | 16 | 0.83 | 45/24 | 1740 |
| 5* | 68/M | 12 | 2.5 | 51/40 | 1950 |

Table 1: Patient Demographics

## 3. Results:

### 3.1 GeMAPS Features:

Table 2 displays our results for GeMAPS features. For α = 0.05, we identified 6 significant features. Since we only collected recordings from five patients, we thought that our small dataset could hinder significant features. As such, Table 2 also includes features that were below a cutoff threshold of $p = 0.1$, leading to 11 total features presented of the 88 total GeMAPS features.

| GeMAPS Feature ID | DBS OFF (Mean ± SD) | DBS ON (Mean ± SD) | p |
|---|---|---|---|
| $F_0$ Semitone Rising Slope Mean | 330.8 ± 208.8 | 223.2 ± 151.9 | 0.002* |
| $F_0$ Semitone 20th Percentile | 26.6 ± 5.6 | 25.6 ± 6.3 | 0.078 |
| $F_0$ Semitone Rising Slope Variance | 470.2 ± 297.2 | 277.1 ± 270.7 | 0.002* |
| $F_2$ Bandwidth Mean | 830.3 ± 97.6 | 867.9 ± 112.9 | 0.018* |
| Mean Unvoiced Segment Length | 0.25 ± 0.19 | 0.30 ± 0.24 | 0.059 |
| Alpha Ratio Mean | -7.11 ± 13.5 | -9.37 ± 13.0 | 0.062 |
| Hammarberg Index Variance | 0.678 ± 0.575 | 0.550 ± 0.365 | 0.065 |
| Loudness Peaks per Second | 2.52 ± 1.13 | 2.11 ± 0.99 | 0.005* |
| Loudness Mean Falling Slope | 5.35 ± 3.13 | 4.40 ± 2.07 | 0.044* |
| Loudness Variance | 0.74 ± 0.23 | 0.81 ± 0.31 | 0.011* |
| Spectral Slope 500-1500 Hz mean | 0.0056 ± 0.0079 | 0.0038 ± 0.0076 | 0.051 |

Table 2: GeMAPS Features Extracted with $p < 0.1$

### 3.2 Prosodic Features:

No statistically significant differences were found for prosody features, thus we decided to look at the correlation between motor symptom severity via the UPDRS part III scores and the different prosodic features. Correlation coefficients between prosody features and motor scores are displayed on Figure 2. Many prosody features significantly correlated with other prosody features. This confirmed our assumption that many of the prosody features were related, i.e. pause length was strongly correlated with pause percentage. Pause percentage, demonstrated the largest correlation with motor scores, with a correlation coefficient of -0.6. All other prosody features were also negatively correlated with motor symptom severity.

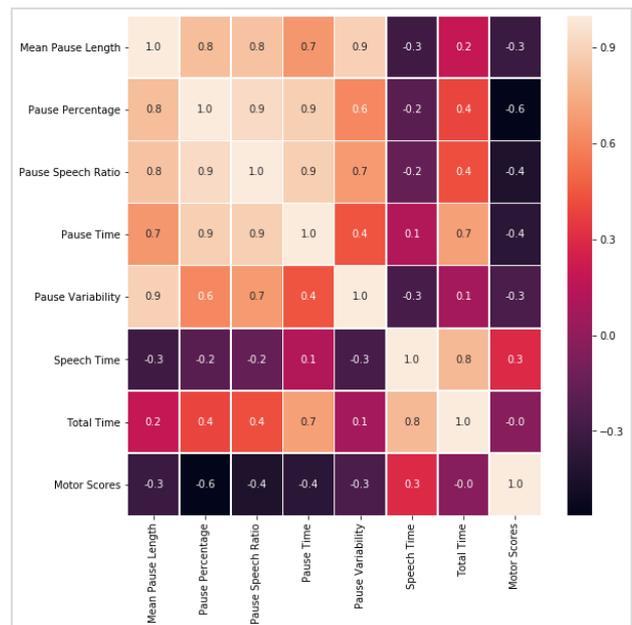

Figure 2: Heatmap for Prosodic Features

### 3.3 Linguistic Features:

Table 3 displays our results for various linguistic features. From the NLTK set we looked at the number of repeats in a task. From Linguistic Features of Complexity, we looked at Brunet's index, Honore's statistic, and the type-token ratio [16]. Differences between these features for DBS ON and OFF were insignificant.

| Linguistic Feature ID | DBS OFF (Mean ± SD) | DBS ON (Mean ± SD) | p |
|---|---|---|---|
| Repeats | 8.6 ± 9.6 | 5.3 ± 5.7 | 0.10 |
| Brunet Index | 7.32 ± 1.90 | 7.15 ± 1.66 | 0.55 |
| Honore Statistic | 481.8 ± 259.0 | 421.2 ± 295.1 | 0.66 |
| Type Token | 0.86 ± 0.12 | 0.90 ± 0.08 | 0.17 |

Table 3: Linguistic Features Extracted

## 4. Discussion:

The purpose of our work was to explore and identify voice derived digital biomarkers, which could potentially help optimize the treatment of PD with DBS. Our identification of significant differences amongst GeMAPS features provides promising results that acoustic voice changes are present and quantifiable. PD has been known to affect many of the motor mechanisms of voice, thus these findings are consistent with PD affecting acoustic properties of voice.

Although none of the prosody features were statistically different for DBS ON vs OFF, negative correlations were present between prosodic features and motor symptom severity. Combined with the correlations that we generated between prosodic features, this is a promising sign that there could be a possible correlation to be further explored. Decreased total electrical energy delivered via DBS has been associated with improvement in speech prosody per [10]. Thus, the correlation we identified is consistent with prior findings, as prosodic features were found to improve with DBS OFF.

Linguistic features explored also demonstrated no statistical difference for DBS ON vs OFF. Although the findings were insignificant, they provide results that could be further explored with an expanded data set. For the complex linguistic features, results seemed to conflict. DBS ON corresponded to a lower Brunet Index, meaning a richer lexical text as compared to the DBS OFF group. As for the Honore statistic, DBS ON was associated with a lower Honore Statistic, a less strong overall lexicon as compared to the DBS OFF group. These results demonstrate the differing linguistic features that could be associated with DBS ON or OFF, but further studies will need to be explored to demonstrate if the relationship is significant.

### 4.1 Limitations:

Our current study was limited by a few factors. The sample size of our study was only five individuals, with an age range of 34 years. This sample presents potential confounding due to varying age-related co-morbidities. Medication dosages were not controlled for, and all patients had differing LED's at the time of data collection which could contribute to measured motor scores and voice characteristics. Lastly, the nature of individual variability on DBS settings for STN/GPI targeted DBS, would generate varying effects in each patient. The small sample size was not able to normalize any of these factors.

## 5. Conclusion:

This study provides preliminary evidence for voice biomarkers to act as an indicator for motor symptom severity amongst PD patients with DBS. Acoustic features provided significant differences associated with DBS in ON and OFF states, as well as prosodic features demonstrating a correlation with motor symptom severity. With the increased accessibility of voice recording devices and voice technology, voice data has potential utility in monitoring and optimizing DBS treatment without the need for patients to come into clinic. Our hope is to be able to gather data amongst a larger subset of DBS patients. This will allow us to utilize a data set where we can strengthen correspondence between motor symptom severity with voice features, and potentially apply towards voice operated closed-loop DBS.

## 6. Statements:

*6.1 Acknowledgements:*

We would like to thank Avery Pong for providing research support during the preliminary phase of the project.

*6.2 Statement of Ethics:*

Ethical oversight for this study was provided by the University of Washington Institutional Review Board (UW IRB study 0790). Subjects who participated were required to sign for informed consent.

*6.3 Disclosure Statement:*

Dr. Hosseini Ghomi is a stock holder of Neurolex Laboratories.

*6.4 Funding Sources:*

Dr. Hosseini Ghomi's work was supported by the VA Advanced Fellowship in Parkinson's Disease, which provided protected research time.

*6.5 Author Contributions:*

R.H.G was the principal investigator, wrote and edited the manuscript, and guided the overall project. H.P carried out participant recruitment, data collection, and significant portions of manuscript writing. S.J carried out data processing, analysis, and portions of manuscript writing. L.Z was a primary research mentor and provided editorial guidance. All authors read and approved the final manuscript.